\documentclass[pra, twocolumn,showpacs,amsmath,amssymb]{revtex4}

\usepackage{graphicx}
\usepackage{dcolumn}
\usepackage{bm}

\begin{document}

\title{Electric Quadrupole Moments of Metastable States of Ca$^+$, Sr$^+$, and Ba$^+$}

\author{Dansha Jiang}
\affiliation{Department of Physics and Astronomy, University of
Delaware, Newark, DE 19716-2570, USA}
\author {Bindiya Arora}
\affiliation{Department of Physics and Astronomy, University of
Delaware, Newark, DE 19716-2570, USA}

\author{M. S. Safronova}
 \homepage{http://www.udel.edu/~msafrono}
 \email{msafrono@udel.edu}
\affiliation{Physique des Interactions Ioniques et Mol\'{e}culaires
(CNRS UMR 6633), Universit\'{e} de Provence, Centre de
Saint-J\'{e}r\^{o}me, Case C21, F13 397 Marseille Cedex 20, France}
\affiliation{Department of Physics and Astronomy, University of
Delaware, Newark, DE 19716-2570, USA\footnote{Permanent address}}

\begin{abstract}
Electric quadrupole moments of the metastable $nd_{3/2}$ and
$nd_{5/2}$ states of Ca$^+$, Sr$^+$, and Ba$^+$ are calculated using
the relativistic all-order method including all single, double, and
partial triple excitations of the Dirac-Hartree-Fock wave function
to provide recommended values for the cases where no experimental
data are available. The contributions of all non-linear single and
double terms are also calculated for the case of Ca$^+$ for
comparison of our approach with the CCSD(T) results. The third-order
many body perturbation theory  is used to evaluate contributions of
high partial waves and the Breit interaction. The remaining omitted
correlation corrections are estimated as well. Extensive study of
the uncertainty of our calculations is carried out to establish
accuracy of our recommended values to be 0.5\% - 1\% depending on
the particular ion. Comprehensive comparison of our results with
other theoretical values and experiment is carried out. Our result
for the quadrupole moment of the $3d_{5/2}$ state of Ca$^+$ ion,
1.849(17)~$ea_0^2$, is in agreement with the most precise recent
measurement 1.83(1)~$ea_0^2$ by Roos et al. [Nature 443, 316
(2006)].

\end{abstract}

\pacs{31.15.A-, 31.15.bw, 32.10.Dk, 06.30.Ft}
\maketitle

\section{Introduction}
Frequency standards based on optical transitions of trapped ions
have the potential to reach a systematic fractional uncertainty on
the order of $10^{-18}$~\cite{Optical-clock}.  The ability to
develop more precise optical frequency standards will open ways to
improve Global Positioning System (GPS) measurements and tracking of
deep-space probes, perform more accurate measurements of the
physical constants and tests of fundamental physics such as searches
for nonlinearity of quantum mechanics, gravitational waves, etc.
  Some of the promising candidates for such ultra-high-precision frequency standards with trapped ions  are
 $^{27}$Al$^+$~\cite{Al,Al1}, $^{199}$Hg$^+$~\cite{New-Hg1, New-Hg2},
  $^{171}$Yb$^+$~\cite{New-Yb1, New-Yb2}, $^{87}$Sr$^+$~\cite{New-Sr1, New-Sr2},
  $^{43}$Ca$^+$~\cite{Prop-Ca},  $^{115}$In$^+$~\cite{In},  and $^{137}$Ba$^+$.
   One of the largest
   sources  of systematic errors in such frequency
   standards with monovalent ions is due to interaction of the quadrupole moments of metastable
    states with stray electric field gradients \cite{Dehmelt,Err-Hg}.
The electric quadrupole moments of the metastable states  are hard
to calculate accurately even for simplest monovalent systems owing
to large correlation corrections (over 30\% for Ca$^+$).
Relativistic configuration interaction (RCI) method with a
multiconfiguration Dirac-Fock orbital basis was used by
Itano~\cite{I06} to calculate relevant quadrupole moments in Ca$^+$,
Sr$^+$, Ba$^+$, Yb$^+$, Hg$^+$, and Au. The RCI results agreed with
available measurement within 10\%. The relativistic coupled-cluster
calculations  of quadrupole moments of metastable $nd$ states were
carried out by Sur \emph{et al}.~\cite{SLSCDM06}  for Ca$^+$,
Sr$^+$, and Ba$^+$ and by Sahoo~\cite{S06Ba} for Ba$^+$. These
calculations yielded results $5\%$ and $13\%$ higher than the recent
measurements of Ca$^+$ \cite{RCKRB06}  and Sr$^+$ \cite{BMHGK04}
quadrupole moments, respectively.
 Mitroy and Zhang~\cite{MZ08, MZB08}  calculated the quadrupole moments of the $3d_{5/2}$ state in Ca$^+$
 and the
 $4d_{5/2}$ state in Sr$^+$ by diagonalizing a semiempirical Hamiltonian in a large dimension single electron
 basis. Their values are in good agreement with the experiment.
 However, they have noted that their particular definition of the polarization potential  may lead to a possible
  problem with the accuracy of properties of these $nd$ states calculated using this method~\cite{MZ08,MZB08}. Large
  differences between theoretical calculations and experimental values,  especially the 5\% discrepancy
   between the recent
   precise measurement of the $3d_{5/2}$ state quadrupole moment of Ca$^+$ by
   Roos \emph{et al.}~\cite{RCKRB06} and accurate
   coupled-cluster single-double partial triple [CCSD(T)] value from \cite{SLSCDM06},
   and consecutive need for comprehensive analysis of the theoretical
   uncertainties have in part motivated this work.

 In this paper, we present the relativistic all-order calculations
 of the
electric quadrupole moments of the $nd_{3/2}$ and $nd_{5/2}$ states
of Ca$^+$, Sr$^+$, and Ba$^+$ ions. The relativistic all-order
method is  one of the most accurate methods used for the calculation
 of atomic properties of monovalent systems (see Ref.~\cite{Advances} for
  a review and references therein).
The lifetimes of the $3d$ levels in Ca$^+$ calculated in this
approach and estimated to be accurate to 1\% were found to be in
agreement with the high-precision experiment \cite{Ca}. The
calculation of the $nd$ quadrupole moments is very similar to the
calculation of the $nd$ lifetimes so similar accuracy is expected.
The long lifetimes of the metastable $nd$ states of these ions also
make these systems well suited for the study of quantum information
processing and quantum simulation \cite{quantum1,quantum2}.

 The atomic properties of Ba$^{+}$ are also of particular interest
owing to the prospects of studying the parity nonconservation  with
a single trapped ion \cite{fortson:93}.
Progress on the related spectroscopy with a single Ba$^+$ ion is 
reported in \cite{koerber:02,koerber:03}, and precision measurements
of light shifts in a single trapped Ba$^+$ ion have been reported in
\cite{Ba:2005}.

  Another motivation for this work is an opportunity to evaluate the
importance of the non-linear terms as well as triple and higher
excitations in the coupled-cluster approach. It has been indicated
in Refs.~\cite{Andrei1,Andrei2} that non-linear terms may be
relatively large and significantly cancel with triple and
higher-excitation terms that are not included in the CCSD(T) of
Ref.~\cite{SLSCDM06} or all-order single-double partial triple
(SDpT) approaches \cite{Advances}. In this work, we  include all
non-linear terms at single-double (SD) level and evaluate triple and
higher-excitation corrections beyond CCSD(T) or SDpT treatments for
Ca$^+$. Our results demonstrate significant cancelation between
these terms. We note that it is the first time that such calculation
has been carried out for any of the atomic properties. Our results
in part explain the discrepancy of previous high-precision
calculation with the Ca$^+$ experiment and represent the most
complete calculation to date. We also present detailed analysis of
the uncertainty of our calculations.

\section{Method}
\label{method}

The electric quadrupole moment $\Theta(\gamma J)$ of an atom in
electronic state $| \gamma J \rangle$ is defined as the diagonal
matrix element of the $q=0$ component of the electric quadrupole
operator $Q$ in a spherical basis
\begin{equation}
\Theta(\gamma J)= \left<\Psi( \gamma JM_J) \left|  Q_0 \right| \Psi(
\gamma J M_J)\right>, \label{eqq}
\end{equation}
with the magnetic quantum number $M_J$ taken to be equal to its
maximum value, $M_J=J$ \cite{I06}. Applying the Wigner-Eckart
theorem and using analytical expression for the relevant 3-j
coefficient \cite{Walter}, allows to express the quadrupole moment
via the reduced matrix element of the quadrupole operator as
\begin{equation}
\Theta(\gamma J)=\frac{(2J)!}{\sqrt{(2J-2)!(2J+3)!}}
 \left<\Psi( \gamma J) \left\|  Q \right\|
 \Psi( \gamma J )\right>,
\end{equation}
where electric quadrupole operator $Q$ is represented in second
quantization as a one-body operator
\begin{equation}
Q =\sum_{ij} q_{ij} a^{\dagger}_i a_j.
\end{equation}
Here, $a^{\dagger}_i$ and $a_i$ are the creation and annihilation
operators.

 In the coupled-cluster method, the exact many-body wave function $\Psi (\gamma J)$
is represented in the form \cite{CK:60}
\begin{equation}
|\Psi \rangle = \exp(S) |\Psi^{(0)}\rangle,  \label{cc}
\end{equation}
where $|\Psi^{(0)}\rangle$ is the lowest-order wave function. We
have omitted indexes ($\gamma J$) in this equation and formulas
below for convenience. The operator $S$ for an N-electron atom
consists of ``cluster'' contributions from one-electron,
two-electron, $\cdots$, N-electron excitations of the lowest-order
wave function $|\Psi^{(0)}\rangle$:
\begin{equation}
S=S_1+S_2+ \dots +S_N.
\end{equation}
The expansion of the exponential in Eq.~(\ref{cc}) in terms of the
$n$-body excitations $S_n$ gives
\begin{equation}
|\Psi \rangle = \left\{ 1+S_1+S_2+\frac{1}{2} S_1^2 + S_1 S_2
+\frac{1}{2} S_2^2+ \cdots \right\} |\Psi^{(0)}\rangle.
\end{equation}
In the linearized coupled-cluster method, only linear terms are
considered, and the wave function takes the form
\begin{equation}
|\Psi \rangle = \left\{1+S_1+S_2+S_3 + \cdots +S_N \right\}
|\Psi^{(0)}\rangle\, .
\end{equation}
We note that the contributions from the nonlinear terms are expected
to be relatively small, but the computational complexity and time
increases significantly with their addition in the present approach
\cite{NL}.
 The relativistic all-order single-double (SD) method is the
linearized coupled-cluster method restricted to single and double
excitations only, with the wave function given by
\begin{eqnarray}
\label{SDeq} &&|\Psi_{\textrm{SD}} \rangle = \left\{1+S_1+S_2
\right\} |\Psi^{(0)}\rangle\
\\ && = \left[ 1+\sum_{ma} \rho_{ma} a^{\dagger}_m a_a
+ \sum_{mv} \rho_{mv} a^{\dagger}_m a_v \right. \nonumber \\
 && \left. + \sum_{mnab} \rho_{mnab} a^{\dagger}_m a^{\dagger}_n a_b a_a
     + \sum_{mna} \rho_{mnva} a^{\dagger}_m a^{\dagger}_n a_a a_v  \right] \vert \Psi^{(0)} \rangle\nonumber,
\end{eqnarray}
where we take frozen-core Dirac-Hartree-Fock (DHF) wave function to
be the lowest-order wave function $\vert \Psi^{(0)} \rangle$.
 The indices $m$ and $n$  designate excited states while indices $a$ and $b$ designate core
  states; the index $v$ labels the valence electron.
The equations for the single excitation coefficients $\rho_{ma}$,
$\rho_{mv}$, double excitation coefficients $\rho_{mnab}$,
$\rho_{mnva}$, and the corresponding  correlation core and valence
energies $\delta E_{core}$, $\delta E_v$
   are solved iteratively in a finite basis set.
The finite basis set used in our calculations
   consists of single-particle orbitals which are linear combinations of B-splines~\cite{Bspline}
   constrained to a spherical cavity.

The all-order SDpT method is an extension of the SD method in which
valence part of the linear triple excitation term $S_3$ is added to
the wave function:
\begin{eqnarray}
|\Psi_{\textrm{SDpT}} \rangle &=&|\Psi_{\textrm{SD}} \rangle \nonumber \\
&+ &\sum_{mnrab} \rho_{mnrvab} a^{\dagger}_m a^{\dagger}_n
a^{\dagger}_r a_b a_a a_v
 \vert \Psi^{(0)} \rangle,
 \label{SDpTeq}
\end{eqnarray}
where  the $|\Psi_{\textrm{SD}} \rangle$ is given by
Eq.~(\ref{SDeq}). The dominant part of $S_3$ is treated
perturbatively, i.e. its effect on the valence energies $\delta E_v$
and single excitation coefficients $\rho_{mv}$ is calculated, but
the equations for  the triple excitation coefficients
$\rho_{mnrvab}$ are not iterated. A detailed description of the SD
and SDpT methods is given in Refs.~\cite{Advances,SD,SDpT}.

We carry out both SD and SDpT calculations in this work to establish
the size of the triple corrections in the perturbative approach. The
coupled-cluster CCSD(T) method used in calculation of the quadrupole
moments in Refs.~\cite{SLSCDM06,S06Ba} also includes the triple
excitations perturbatively even though  the particular terms that
are considered may somewhat differ.

 In this work, we also carry out the all-order calculation that
 includes all non-linear terms that arise from single and double
 excitation terms  $S_1$ and $S_2$ for the case of Ca$^+$. There are only
 six of such terms that can contribute to
 the equations for single and double excitation coefficients, and the
 complete coupled-cluster single-double (CCSD) wave function is then written as
\begin{eqnarray}
\label{CCSDeq} &&|\Psi_{\textrm{CCSD}} \rangle = \exp(S_1+S_2)
|\Psi^{(0)}\rangle =|\Psi_{\textrm{SD}} \rangle  + \\ &&\left\{
\frac{1}{2}S^2_1+S_1S_2+\frac{1}{2}S^2_2+\frac{1}{6}S^3_1
+\frac{1}{2}S^2_1S_2+\frac{1}{24}S^4_1   \right\}
|\Psi^{(0)}\rangle, \nonumber
\end{eqnarray}
where  the $|\Psi_{\textrm{SD}} \rangle$ is given by
Eq.~(\ref{SDeq}). The complete formulas for the CCSD equations are
given in Ref.~\cite{NL}. Our approach allows us to explicitly
calculate the contribution of the non-linear terms to the quadrupole
moments as the difference of the results obtained in the CCSD and SD
approaches.

  The matrix element of any one-body operator Z in the all-order
  method
 is obtained as
\begin{equation}
 Z_{vw} = \frac{\langle \Psi_v \vert Z \vert \Psi_w
\rangle } {\sqrt{\langle \Psi_v \vert \Psi_v \rangle \langle \Psi_w
\vert \Psi_w \rangle}}.
\end{equation}
For non-scalar operators, this expression becomes
\begin{equation}
 Z_{vw}=\frac {Z_{\rm val}} {\sqrt{(1+N_v)(1+N_w)}}, \label{matrix}
\end{equation}
 where
 the expression for
the numerator of Eq.~(\ref{matrix}) derived with
$|\Psi_{\textrm{SD}}\rangle$ wave function consists of the sum of
the DHF matrix element $z_{wv}$ and twenty other terms $Z^{(k)}$,
$k=a\cdots t$. These terms and the normalization terms $N_v$  are
linear or quadratic functions of the excitation coefficients
$\rho_{ma}$, $\rho_{mv}$, $\rho_{mnab}$, and $\rho_{mnva}$.
 The
complete expression for the matrix elements can be found in
\cite{MBPT}. The same expression [Eq.(\ref{matrix})] for the matrix
elements is used in all calculations in this work.

We carry out three different \textit{ab initio} calculations of the
quadrupole matrix elements. In the first one, all excitation
coefficients are obtained in the SD approach [Eq.~(\ref{SDeq})], in
the second one they are obtained in the SDpT approach
[Eq.~(\ref{SDpTeq})], and in the third one (carried out for Ca$^+$)
the excitation coefficients are obtained in the CCSD approach
[Eq.(\ref{CCSDeq})]. We refer to these results as SD, SDpT, and CCSD
values in the text and tables below.

While the numerator of the Eq.~(\ref{matrix}) contains twenty
correlation terms, only one term is overwhelmingly dominant for the
quadrupole moments considered in this work, contributing over 90\%
of the total correlation correction. Following the notation of
Ref.~\cite{MBPT}, this is the term $Z^{(c)}$ that is equal in the
case of the diagonal quadrupole matrix element to

\begin{table*}
\caption{\label{tab1} Contributions of high partial waves and Breit
interaction to the  electric quadrupole moments of Ca$^+$ calculated
using third-order many-body perturbation theory; $l_{max}$ is the
highest number of partial waves included in the particular
calculation.  All values are given in atomic units.}
\begin{ruledtabular}
\begin{tabular}{lcccccc}
 State  &$l_{max} = 6$&  $l_{max} = 8$& $l_{max} = 10$& $l_{max} = 12$&$l = 7\dots12$ & Breit\\
 \hline
$3d_{3/2}$ &  1.134 &   1.127     &   1.124     & 1.123 &-0.011
&-0.001\\
$3d_{5/2}$&  1.628&  1.617&   1.614&   1.612&   -0.015&  -0.003\\
\end{tabular}
\end{ruledtabular}
\end{table*}

\begin{equation}
Z^{(c)} = 2\sum_{m} q_{vm} \rho_{mv}, \label{termc}
\end{equation}
where sum over $m$  ranges over all excited basis set states. The
lowest-order DHF matrix elements $q_{ij}$ of the quadrupole operator
are given by
\begin{equation}
q_{ij}=\langle i\|C^{(2)}\| j\rangle \int^{\infty}_0 r^2\ \left(
g_i(r)g_j(r) + f_i(r)f_j(r)\right) dr,
\end{equation}
where $C^{(2)}$ is the normalized spherical harmonic of rank 2 and
$g_i$, $f_i$ are large and small components of the Dirac wave
function, respectively. The $\rho_{mv}$ are single valence
excitation coefficients calculated in either SD [Eq.~(\ref{SDeq})],
SDpT [Eq.~(\ref{SDpTeq})], or CCSD [Eq.~(\ref{CCSDeq})]
approximations as described above. Therefore, evaluation of the
omitted higher-order corrections to $\rho_{mv}$ provides an estimate
of the dominant part of the missing contributions in each
approximation. These excitation coefficients are closely related to
the correlation energy $\delta E_v$. If we introduce the self-energy
operator (also referred to as correlation potential in some works)
$\Sigma_{mv}$ as
\begin{equation}
\Sigma_{mv}=\left( \epsilon_v - \epsilon_m +\delta E_v \right)
\rho_{mv},
\end{equation}
where $\epsilon_i$ is the DHF energy of the state $i$, then the
correlation energy would correspond to the diagonal term
$\Sigma_{vv}$. Therefore, the omitted correlation correction can be
estimated by adjusting the single-excitation coefficients
$\rho_{mv}$ to the experimentally well-known value of the valence
correlation energy, and then re-calculating the matrix elements
using Eq.~(\ref{matrix}) with the modified coefficients \cite{SD}
\begin{equation}
\rho_{mv}^{\prime}=\rho_{mv} \frac{\delta
E_v^{\textrm{expt}}}{\delta E_v^{\textrm{theory}}}. \label{scale}
\end{equation}
 The $\delta E_v^{\textrm{expt}}$ is defined as the experimental
 energy \cite{NIST}
minus the lowest order DHF energy $\epsilon_v$. The theoretical
correlation energy is somewhat different in the SD, SDpT, and CCSD
approaches. Therefore, this scaling procedure has to be conducted
separately for each of these three calculations with $\delta
E_v^{\textrm{theory}}$ taken to be $\delta E_v^{\textrm{SD}}$,
$\delta E_v^{\textrm{SDpT}}$, and $\delta E_v^{\textrm{CCSD}}$,
respectively. We refer to the results of these calculations as
SD$_{sc}$, SDpT$_{sc}$, and CCSD$_{sc}$ values.

Before discussing the final results of our calculations, we describe
the calculation of two other corrections that need to be accounted
for in the \textit{ab initio} SD, SDpT, and CCSD calculations. Any
sum over the excited states in either the calculation of the
excitation coefficients or matrix elements using the
Eq.~(\ref{matrix}) involve the sum over the principal quantum
number, calculated essentially exactly, and the sum over the partial
waves, that needs to be truncated after some value $l_{max}$ (see
the sum over $m$ in Eq.~(\ref{termc}) for an example). In all of our
all-order calculations, we chose $l_{max}=6$. We find that the
contributions from higher partial waves are small but significant
and should not be omitted at the present level of accuracy. The size
of the contribution of the higher partial waves may also shed some
light on the disagreement of some previous calculations with
experiment.

 To evaluate this
contribution, we first carried out a
  third-order many-body perturbation theory (MBPT) calculation with the same basis
   set and $l_{max}$ as the all-order
  calculations, and
  then performed the same calculation with larger basis set and  larger
  $l_{max}$.
  A detailed description of the third-order MBPT method is given in Ref.~\cite{3rd-MBPT}.

The results of the third-order calculation with the increasing
values of $l_{max}$ for the quadrupole moments of Ca$^+$ are given
in Table~\ref{tab1}. While the total contribution of the $l=7, 8$
partial waves is rather substantial, 0.6\%, contributions of even
higher partial waves are small. We truncated the sum after $l = 12$,
with the expected uncertainty of this truncation being well below
our final accuracy. The difference between
   the MBPT calculation with $l_{max} = 6$ and $l_{max} = 12$ is taken to be the
   correction for the contribution of higher partial waves   and is added to the
   \textit{ab initio} all-order results.

We have also evaluated the total contribution of the $l = 5, 6$
partial waves to establish its size, and found it to be 3\%.
Moreover, the inclusion of larger number of partial waves
\textit{reduces} the values of the quadruple moments since term
$Z^{(c)}$ contributes with the sign opposite to that of the
lowest-order value. The inclusion of larger number of partial waves
increases the absolute value of the correlation correction leading
to lower total final values. Therefore, omitting contributions of
higher partial waves or exclusion of such orbitals from the basis
set in other calculations may result in an overestimation of the
quadrupole moments by a few percent.

 We also investigated the effect of
the Breit interaction which arises due to exchange of a virtual
photon between atomic electrons and can be written as
\begin{equation}\label{Breit}
 B_{ij}= - \frac {1}{r_{ij}} \hspace{1mm}
\mbox{\boldmath $\alpha$}_i \cdot \mbox{\boldmath $\alpha$}_j +
\frac{1}{2r_{ij}} \left[\hspace{1mm} \mbox{\boldmath $\alpha$}_i
\cdot \mbox{\boldmath $\alpha$}_j - \left( \mbox{\boldmath
$\alpha$}_i \cdot \mathbf{\hat{r}}_{ij} \right)
\left(\mbox{\boldmath $\alpha$}_j \cdot \mathbf{\hat{r}}_{ij}
\right)\right],
\end{equation}
where \mbox{\boldmath $\alpha_i$} are the Dirac matrices. This
correction includes instantaneous magnetic interaction between Dirac
currents (the first term) and the retardation correction to the
electric interaction (the second term). In order to calculate the
Breit correction to the quadrupole matrix elements, we modify the
generation of B-spline basis set to intrinsically include the Breit
interaction on the same footing as the Coulomb interaction, and
repeat the third-order calculation with the
 modified basis set. The difference between the new values and the original third-order calculation
 is taken to be the correction due to Breit interaction. This contribution is listed in the last column of
 Table~\ref{tab1}.
\begin{table*}
\caption{\label{tab2}Electric-quadrupole moments of Ca$^+$, Sr$^+$,
and Ba$^+$ calculated using
 different approximations: Dirac-Hartree-Fock (DHF),  third-order many-body perturbation theory (MBPT),
  single-double all-order method (SD), and single-double all-order method including partial
   triple excitation contributions (SDpT), label "sc" represent the corresponding scaled values.
  The results of the full single-double couple-cluster calculation for Ca$^{+}$ are listed in row
  labeled ``CCSD''; the corresponding scaled values are listed in the row ``CCSD$_{sc}$''.
  All values are given in atomic units.}
\begin{ruledtabular}
\begin{tabular}{llccccccccc}
Ion &State&     DHF&    MBPT&   SD& SDpT&   SD$_{sc}$&SDpT$_{sc}$&    CCSD &CCSD$_{sc}$&  Final\\
\hline Ca$^+$  &$3d_{3/2}$&    1.712&  1.122&  1.245&  1.282& 1.289&
1.281& 1.271  & 1.292 &  1.289(11)\\
&$3d_{5/2}$&            2.451&  1.610&  1.785&  1.837&  1.849&
1.836& 1.822   &1.851
&  1.849(17)\\[0.3pt]
Sr$^+$  &$4d_{3/2}$&    2.469&  1.876&  1.987&  2.021&  2.029&  2.020& & & 2.029(12)\\
&$4d_{5/2}$&            3.559&  2.721&  2.876&  2.922&  2.935&  2.923& & & 2.935(17)\\[0.3pt]
Ba$^+$  &$5d_{3/2}$&    2.732&  2.086&  2.217&  2.260&  2.256&  2.248& & & 2.256(11)\\
&$5d_{5/2}$&            3.994&  3.087&  3.263&  3.323&  3.319&  3.308& & & 3.319(15)\\
\end{tabular}
\end{ruledtabular}
\end{table*}
  In second quantization, Breit interaction operator in a normal form
  is separated into a one-body part and a two-body part~\cite{Breit}.
  The two-body Breit contribution is omitted in our approach. The total Breit
     corrections are small  and are
   below the estimated uncertainty of our theoretical values.
   Therefore, the possible uncertainty introduced by the omission of
   the two-body Breit correction is negligible. In fact, we find that most of the
   Breit correction arises at the DHF level.

     \section{Results and Discussion}

\begin{table*}[htpb]
\caption{\label{tab3} Comparison of the present results for electric
quadrupole moments in Ca$^+$, Sr$^+$, and Ba$^+$
 with other calculations and experiments. All values are in atomic units.}
\begin{ruledtabular}
\begin{tabular}{lccccccc}
Ion & State & Present  &Ref.~\cite{I06}  & Ref.~\cite{SLSCDM06} & Other & Expt.\\
\hline
Ca$^+$ & $3d_{3/2}$&1.289(11)&1.338& 1.338&\\
&  $3d_{5/2}$ & 1.849(17)& 1.917&
1.916&1.819\footnote{Reference~\protect~\cite{MZ08}}&1.83(1)\footnote{Reference~\protect~\cite{RCKRB06}}
\\[0.2pc]
Sr$^+$& $4d_{3/2}$&2.029(12)&2.107& 2.12&\\
&  $4d_{5/2}$&2.935(17)&3.048&
2.94(7)&2.840\footnote{Reference~\protect~\cite{MZB08}}&
2.6(3)\footnote{Reference~\cite{BMHGK04}}&\\[0.2pc]
Ba$^+$ &$5d_{3/2}$&2.256(11)&2.297& 2.309&2.315\footnote{Reference~\cite{S06Ba}, CCSD(T)} &\\
&  $5d_{5/2}$&3.319(15)&3.379& 3.382&3.382$^e$\\
\end{tabular}
\end{ruledtabular}
\end{table*}

The results of our calculations of the quadrupole moments of the
metastable $nd_{3/2}$ and $nd_{5/2}$ states of Ca$^+$, Sr$^+$, and
Ba$^+$ ions are summarized in Table~\ref{tab2}, where we list the
lowest order DHF, third-order MBPT, all-order SD and SDpT \textit{ab
initio}, and corresponding all-order scaled
 values calculated as described in Section~\ref{method}.
 In the case of Ca$^+$, we also list the results of our CCSD
 and scaled CCSD calculations. The \textit{ab  initio} values
 contain  the corrections for the higher partial wave contributions
 and Breit interaction. These corrections do not need to be included
 into the scaled results as that will lead to double counting of
 these effects.
We take the scaled SD numbers as the final values based on the
comparisons of similar calculations
 in alkali-metal atoms with experiment (see Refs.~\cite{Rb,US-1,US-2,US-3,US-4}
 and references therein).

  We take the maximum difference between  the final
  values and the SDpT \textit{ab initio}, SDpT scaled [SDpT$_{sc}$], and CCSD scaled [CCSD$_{sc}$]
   values to be the uncertainty
  of the dominant contribution. We assume that any remaining uncertainty does not exceed the
  uncertainty of the dominant term and take it to be equal to the uncertainty in the dominant term
  evaluated as described above.
The two uncertainties are added in quadrature to obtain the final
estimate of the uncertainty of our values.

We make several conclusions  from our results (all the \%  numbers
are given for Ca$^+$, but the general trends are the same for all
ions considered in this work) :
\begin{enumerate}
\item The triple contributions included in the perturbative approach
contribute about 3\% and \textit{increase} the values of the
quadrupole moments.
\item
The non-linear terms contribute about 2\% and also \textit{increase}
the values of the quadrupole moments.
\item While the SD, SDpT, and CCSD results vary by a few percent, the addition of the
estimated omitted correlation correction carried out according to
Eq.~(\ref{scale}), brings all these results to very close agreement
providing additional validation of this procedure.
\item The
linearized SD$_{sc}$ and complete coupled-cluster CCSD$_{sc}$ scaled
results are nearly exactly the same, with the differences being well
below our estimated uncertainty. Therefore, we found it unnecessary
to carry out CCSD calculations for Sr$^+$ and Ba$^+$.
\item We confirm that non-linear terms strongly cancel with the
triple and higher-excitation contributions not included in the
perturbative approach. As a result, CCSD(T) method used in
Refs.~\cite{SLSCDM06,S06Ba}  that includes both non-linear terms and
triple excitations in the perturbative approach only, is expected to
yield results a few percent higher than the experimental values.
\end{enumerate}

In Table~\ref{tab3}, we compare our final values with other
calculations and available experimental results. We note that our
calculation is the most complete one at the present time.

The J-independent moments, i.e. the values with the all angular
factors divided out, can be obtained by multiplying our results in
Table~\ref{tab3} by $5$ and $7/2$ for the $nd_{3/2}$ and $nd_{5/2}$
states, respectively, according to Eq.~(\ref{eqq}).

Our values are systematically lower than the results of relativistic
configuration interaction (RCI) calculation carried out with a
multiconfiguration Dirac-Fock orbital basis by Itano~\cite{I06}. As
we noted above, high partial waves ( $l > 4$) contribute
significantly (about 4\%) to the quadrupole moments and reduce the
values. Therefore, the restriction of the excitations to mostly
low-$l$ orbitals in Ref.~\cite{I06}  is expected to lead to higher
values in RCI calculations.
 The relativistic coupled-cluster CCSD(T) results
by Sur \emph{et al}.~\cite{SLSCDM06} for Ca$^+$, Sr$^+$, and Ba$^+$
and by Sahoo~\cite{S06Ba} for Ba$^+$ are also systematically lower
than our values, with the exception of the $4d_{5/2}$ Sr$^+$
quadrupole moment, which is in agreement with our value. It is
unclear why this one value compares differently. Since this
particular value was the focus of the work \cite{SLSCDM06}, perhaps
it was treated differently from the other cases. As we noted above,
we expect the CCSD(T) results of Refs.~\cite{SLSCDM06,S06Ba} to be a
few percent too high owing to the cancelation of the non-linear
terms and higher-excitation terms not included in CCSD(T) approach.
Another possible issue is the treatment of the high partial wave
contributions. While the tests of various basis sets were conducted
in Ref.~\cite{SLSCDM06}, it is not stated how high partial waves
were considered. We note that the implementation of the
coupled-cluster method in Refs.~\cite{SLSCDM06,S06Ba} is
significantly different from ours and is more closely related to the
quantum chemistry calculations.

The results of Mitroy and Zhang~\cite{MZ08, MZB08} calculated by
diagonalizing a semiempirical Hamiltonian in a large dimension
single electron
 basis are in good agreement with experiment. Our analysis of the
 correlation correction is consistent with such results.
 We demonstrated in Section~\ref{method} that the dominant part of the correlation correction
 to quadrupole matrix elements comes from the term containing essentially the
 correlation potential $\Sigma_{vm}$ that is closely related to the correlation
  energy. Since the cutoff function in the
 semiempirical potential used in Ref.~\cite{MZ08, MZB08} is adjusted
   to reproduce experimental binding energies, it appears to be
   a good representation for this application.

 Our result
for the $3d_{5/2}$ Ca$^+$ quadrupole moment 1.849(17)~$ea_0^2$
 agrees within the quoted
uncertainties with the recent high-precision measurement
1.83(1)~$ea_0^2$  reported by Roos \emph{et al.} in
Ref.~\cite{RCKRB06} that was carried out using a decoherence-free
subspace with specially designed entangled states of trapped ions.
We also verify (by varying the nuclear parameter) that our value is
not depended on the particular isotope within our accuracy. Our
result for the Sr$^+$ $4d_{5/2}$ quadrupole moment
2.935(17)~$ea_0^2$
 is in good agreement (just outside of
the upper $1\sigma$ bound) with experimental value 2.6(3)~$ea_0^2$
by Barwood \emph{et al.}~\cite{BMHGK04}  measured with a single
laser-cooled ion
 confined in an end cap trap with variable dc quadrupole potential.

\section{Conclusion}

In summary, we performed a relativistic coupled cluster calculation of the electric quadrupole
 moments for the $nd_{3/2}$ and $nd_{5/2}$ states of Ca$^+$, Sr$^+$, and Ba$^+$ ions.
Our analysis of various contributions in part explains the
discrepancy of previous high-precision theory with experiment. We
also present detailed evaluation of the uncertainty of our results
and provide recommended values for the cases where no precision
experiments are available. Our result for the quadrupole moment of
the $3d_{5/2}$ state of Ca$^+$ ion 1.849(17)~$ea_0^2$ is in
agreement with the recent measurement 1.83(1)~$ea_0^2$ by Roos et
al.~\cite{RCKRB06}.

\section{Acknowledgement}

M. S. Safronova would like to thank Fernande Vedel, Caroline
Champenois, and Martina Knoop for interesting discussions of the
Ca$^{+}$ frequency standard as well as to thank Physique des
Interactions Ioniques et Mol\'{e}culaires, CNRS - Universit\'{e} de
Provence for hospitality and financial support. This work was
supported  in part by the U.S.A. National Science Foundation Grant
No.\ PHY-0758088.


\end{document}